\documentclass[final]{ustcstep}

\usepackage{color}
\usepackage{abstract}
\usepackage{times}
\usepackage{multicol}
\usepackage{graphics}
\usepackage[figuresleft]{rotating}
\usepackage{lscape}
\usepackage{bbding}
\usepackage{amssymb}
\usepackage{array}
\usepackage{textcmds}
\usepackage{float}
\usepackage{ulem}
\usepackage[square]{natbib}

\def\gsim{\;\lower4pt\hbox{${\buildrel\displaystyle >\over\sim}$}\;}
\def\lsim{\;\lower4pt\hbox{${\buildrel\displaystyle <\over\sim}$}\;}
\def\grls{\;\lower4pt\hbox{${\buildrel\displaystyle >\over <}$}\;}

\newcommand\addr[2]{{\footnotesize \it $^{#1}$#2}\\}

\def\apj{ApJ}
\def\apjl{ApJL}

\def\solphys{Solar Phys.}
\def\aap{A\&A}
\def\pasa{PASA}

\begin{document}

\title{On the Magnetic and Energy Characteristics of Recurrent Homologous Jets from An Emerging Flux}

\author{Jiajia Liu$^{1,4,*}$, Yuming Wang$^{1,5}$, Robertus Erd{\'e}lyi$^{2}$, Rui Liu$^{1,5,6}$, Scott W. McIntosh$^{3}$, Tingyu Gou$^{1,4}$,\\
Jun Chen$^{1,4}$, Kai Liu$^{1,4}$, Lijuan Liu$^{1,4}$ and Zonghao Pan$^{1,4}$\\[1pt]
\addr{1}{CAS Key Laboratory of Geospace Environment, Department of Geophysics and Planetary Sciences, University of Science and Technology of China,  Hefei 230026, China}
\addr{2}{Solar Physics and Space Plasma Research Center (SP2RC), School of Mathematics and Statistics, The University of Sheffield, Sheffield S3 7RH, UK}
\addr{3}{High Altitude Observatory, National Center for Atmospheric Research, P.O. Box 3000, Boulder, CO 80307, USA}
\addr{4}{Collaborative Innovation Center of Astronautical Science and Technology, Hefei 230026, China}
\addr{5}{Synergetic Innovation Center of Quantum Information and Quantum Physics, University of Science and Technology of China, Hefei 230026, China}
\addr{6}{Mengcheng National Geophysical Observatory, University of Science and Technology of China, Hefei 230026, China}
\addr{*}{Corresponding Author, Contact: ljj128@ustc.edu.cn}}

\maketitle
\tableofcontents

\begin{abstract}
In this paper, we present the detailed analysis of recurrent homologous jets originating 
from an emerging negative magnetic flux at the edge of an Active Region.The observed 
jets show multi-thermal features. Their evolution shows high consistence with the characteristic 
parameters of the emerging flux, suggesting that with more free magnetic 
energy, the eruptions tend to be more violent, frequent and blowout-like. The average temperature, average 
electron number density and axial speed are found to be similar for different jets, indicating that 
they should have been formed by plasmas from similar origins. Statistical analysis of the jets and 
their footpoint region conditions reveals a strong positive relationship between the footpoint-region total 
131 \AA\ intensity enhancement and jets' length/width. Stronger linearly positive 
relationships also exist between the total intensity enhancement/thermal energy of the footpoint regions and jets' mass/kinetic/thermal 
energy, with higher cross-correlation coefficients. All the above results, together, confirm the direct relationship between the 
magnetic reconnection and the jets, and validate the important role of magnetic reconnection in transporting
large amount of free magnetic energy into jets. It is also suggested that there should be more free energy released during the 
magnetic reconnection of blowout than of standard jet events.
\end{abstract}

\section{Introduction}

Solar jets are bulks of plasma material ejected along elongated trajectories from 
the solar surface. They are one of the most common dynamic phenomena occuring 
within the solar atmosphere and could be found in active, quiet-Sun and 
polar regions. Based on their spatial scales, jets may be divided into 
two classes: large-scale and small-scale jets. Small-scale jets are ubiquitous and, here, include 
type-I and type-II spicules (the latter also referred to as 
rapid blue excursions, RBEs) in the chromosphere/transition region 
\citep[e.g.,][]{Becker1968, Sterling2000, DePontieu2007, vanderVoort2009, Cranmer2015, Kuridze2015} 
and quasi-periodic intensity perturbations in 
the corona \citep[e.g.,][]{Verwichte2009, Threlfall2013, LiuJ2015b}. The 
importance of small-scale jets are well-known as they are suggested to contribute 
to coronal heating and/or solar wind acceleration \citep[e.g.,][]{Shibata2007, Moore2011, Tian2014}.

Compared with small-scale jets, large-scale jets are more evident, even observable by lower-resolution instruments like 
{\it STEREO}/EUVI and {\it SOHO}/EIT. Based on their different dominant temperatures, 
large-scale jets are sometimes referred to as H$\alpha$ surges with cold plasmas 
\citep[e.g.,][]{Roy1973, Canfield1996}, UV/EUV jets or macrospicules with warm plasmas 
\citep[e.g.,][]{Bohlin1975, LiuJ2014, Bennett2015}, X-ray jets with hot plasmas 
\citep[e.g.,][]{Shibata1992, Cirtain2007} and white-light jets seen in 
white-light coronagraph \citep[e.g.,][]{Moore2015, LiuJ2015, Zheng2016}. The above classification is 
approximate and not absolute, because jets are often found to be formed of multi-thermal 
plasmas and thus observed in multi-passbands. Attributed to observational facts of jets 
including their ``Reverse-Y" shape and the accompanied nano-flares (or brightening), it is now widely believed that large-scale jets are most 
likely to be triggered by magnetic reconnection, especially the interchange reconnection between 
closed and (locally) open magnetic field lines \citep[see, e.g.,][]{Shibata1996, Scullion2009, Pariat2015}.

In simulations, free magnetic energy can be introduced by at least two ways preceding the eruption of jets: flux 
emergence/cancellation \citep[e.g.,][]{Moreno-Insertis2008, Murray2009, Fang2014}, or rotational/shearing motion at the footpoint region 
\citep[e.g.,][]{Pariat2009, Yang2013}. In the first scenario, reconnection happens between the pre-existing 
open flux and newly emerging closed flux. To ensure the onset of magnetic reconnection, the emerging flux 
should contain certain amount of free magnetic energy or a persistent Poynting flow from below 
the photosphere to inject the energy. It is then not surprising recurrent jets might be triggered 
by repeated magnetic reconnection during the emergence of magnetic flux \citep[e.g.][]{Chen2015}. 
However, the question, that how the occurence of recurrent jets 
is influenced by the emerging flux in observations, still needs to be addressed.

As efforts in studying recurrent jets \citep[e.g.,][]{Pariat2010, Chen2015} from observational perspectives, \cite{Guo2013},
\cite{Zhang2014} and \cite{Li2015} studied the photospheric current patterns, successive blobs 
and the quasi-periodic behavior of recurrent jets from newly emerging fluxes at the edge of three different active regions, 
respectively. \cite{Archontis2010} performed a 3D MHD numerical simulation in which a small active region is constructed 
by the emergence of a toroidal magnetic flux tube. As a result of the new emerging flux, successive magnetic 
reconnections set in and a series of recurrent jets erupt. The above studies all support the importance of magnetic 
reconnection in triggering recurrent jets. The natural question arises: how much the reconnection 
influences the properties (such as length, width and mass etc.)
of jets in observations is vital to understanding the related physical processes. Moreover, statistical study of the 
relations between the energies of jets and of the corresponding footpoint regions will help us explore how free magnetic 
energy is distributed during magnetic reconnection.

In this paper, we will perform a statistical analysis of 
recurrent homologous jets from one emerging (negative) flux at the edge of a part with positive magnetic field of an active region. 
Study on the 
evolution of the photospheric magnetic field {\it SDO}/HMI observations is shown in Sect. 2. 
A combined analysis on the jets observed by {\it SDO}/AIA and several characteristic 
magnetic parameters including the photospheric mean current density, the mean current helicity, 
and the total photospheric free magnetic energy is 
performed in Sect. 3 and 4 to investigate the synchronism between the evolution of the jets and their 
magnetic field conditions at footpoints. In Sect. 5 and 6, we present the statistical studies of 
the properties of jets and their corresponding footpoint regions. Our study will be summarized in Sect. 7.

\section{The Photospheric Magnetic Field}

\begin{figure}[tbh!]
\centering
\includegraphics[width=\hsize]{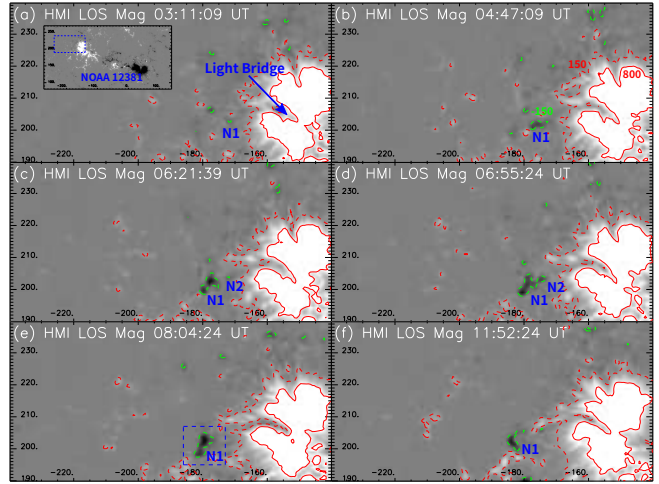}
\caption{Temporal evolution of the emerging negative flux (``N1") and nearby positive polarity (the main 
polarity) of Active Region NOAA 12301 observed by the {\it SDO}/HMI instrument from 
03:00 UT to 12:00 UT. The blue dashed box in the small landscape in panel (a) defines the scope of all the 
panels from (a) to (f). Red solid and dashed curves indicate LOS magnetic field levels at 
800 G and 150 G respective, with the green curves -150 G. Blue dashed box in panel (e) 
covers the region of the emerging negative polarity which results in more than ten times of 
jet eruptions. Coordinates are in units of arc second.}\label{mag}
\end{figure}

Jets studied in this paper are found to be related to an emerging negative polarity at the north-east 
edge of Active Region NOAA 12301 from around 03:00 UT to 12:00 UT on July 9th 2015. 
This Active Region, with a large-scale quadrupolar configuration (for a small landscape in 
Figure~\ref{mag}(a)), turns to the front with a very small positive latitude on late 
July 3rd and is almost at the central meridian during the time window from 03:00 UT to 12:00 UT on July 9th.
Besides the line-of-sight (LOS) magnetic field from {\it SDO}/HMI (Fig.~\ref{mag}(a)), vector magnetic 
field data is also available for this Active Region by the Space-weather HMI 
Active Region Patches \citep[SHARPs,][]{Bobra2014} with SHARP NO. 5745.

Figure~\ref{mag} shows the evolution of the northern part with of the Active Region and the nearby
emerging negative polarities during the aforementioned period. All HMI observations have 
been de-rotated to 00:00 UT.
Solid and dashed red curves in all the panels indicate positive LOS magnetic field with values of 
800 G and 150 G, respectively. 
The dashed curve is employed to show the edge of the Active Region with magnetic strength large enough. 
On the other hand, a tunnel through the 
core of the positive polarity with relatively weak magnetic field is then highlighted by the 
red solid curve, and usually called as the``light bridge" due to its bright appearance compared 
to the sunspot umbra \citep[e.g.,][]{Bray1964, Leka1997, Toriumi2015}.

There are several negative polarities at the edge of the main positive polarity (green 
dashed curves at -150 G in Fig.~\ref{mag}). The one we focus on, labeled as ``N1" in panel (a), becomes the biggest among 
them and finally results in more than ten jet eruptions. The first jet, initiated 
from ``N1", occurs at about 03:11 UT which is only about 6 minutes after its emergence (Fig.~\ref{mag}(a)). 
We notice that before 04:00 UT, the light bridge also enables several jet-like eruptions. The 
relationship between light bridges and jets has been studied \citep[e.g.,][]{LiuS2012}, 
and is beyond the scope of this paper. 

``N1" becomes larger and larger after its emergence, which might indicate continuous flux injection from 
underneath the photosphere. Another evident jet emerges at 04:47 UT (Fig.~\ref{mag}(b)) and 
the size of the emerging negative polarity ``N1" is already much larger than that in panel (a). With 
the growth of ``N1", another negative polarity, ``N2", appears closer to the main positive polarity at 06:21 UT 
(Fig.~\ref{mag}(c)) and merges with ``N1" at around 06:55 UT (Fig.~\ref{mag}(d)). After the 
merging, the eruptions of jets seem to cease, and at 08:04 UT (Fig.~\ref{mag}(e)), a jet much longer than 
all of the previous ones pops out. Similar eruptions with long jets and short time intervals 
last until around 10:10 UT, when the area of ``N1" becomes obviously smaller than it was at 08:00 UT. The 
last jet eruption we study is at around 11:52 UT, when the area of ``N1" already becomes rather small.

\section{Chromopsheric and Coronal Response}

\begin{figure}[tbh!]
\centering
\includegraphics[width=0.95\hsize]{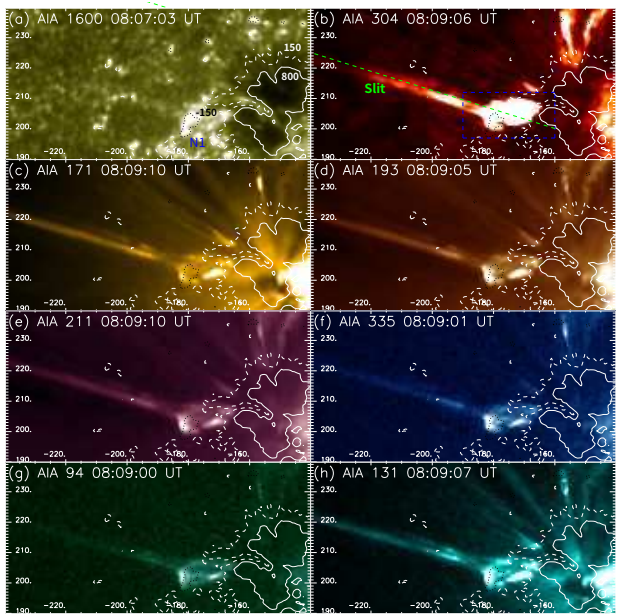}
\caption{{\it SDO}/AIA observations of the sample jet formed at around 08:09 UT at 1600 \AA\ (a), 304 \AA\ (b), 171 \AA\ (c), 
193 \AA\ (d), 211 \AA\ (e), 335 \AA\ (f), 94 \AA\ (g) and 131 \AA\ (h) passbands, respectively. Similar to the red solid, red dashed 
and green dashed contours in Fig.~\ref{mag}, the white solid, white dashed and black dashed contours 
here represent magnetic field levels of 800 G, 150 G and -150 G, respectively. The green   dashed line 
in panel (b) indicate a 13$''$-wide slit for probing the axial motion of jets. The blue dashed box confines 
the region for calculating the normalized integrated intensity in Fig.~\ref{slit}(b). Coordinates 
are in units of arcsecond.}\label{jet}
\end{figure}

There are more than 10 jet eruptions originating from the emerging negative flux ``N1" between 
04:00-12:00 UT, observed by the {\it SDO}/AIA instrument. Most of the jets contain materials that can be 
found in all 7 AIA UV/EUV passbands (see the online movie M1, a supplement to Figure~\ref{jet}), suggesting 
their multi-thermal nature.

Among all these prominent jets, Figure~\ref{jet} captures one at around 08:09 UT from the upper photosphere to the corona. 
Panel (a) shows the upper photospheric observation of the emerging flux and the nearby active region. Contours are 
similar to those in Figure~\ref{mag} from the {\it SDO}/HMI LOS magnetic field data, with white solid, white dashed 
and black dashed curves indicating the LOS magnetic field strength at 800 G, 150 G and -150 G, respectively. 
The black dashed curve, representing the emerging negative flux ``N1", coincides with the brightness enhancement for the footpoint region of the jet, 
providing a direct evidence that the jet is initiated from the emerging flux. 

Figures~\ref{jet}(b)-(h) show the UV/EUV observations of the jet at passbands of He II 304 \AA\ (0.05 MK), Fe IX 171 \AA\ (0.6 MK), 
Fe XII XXIV 193 \AA\ (1.6 MK, 20 MK), Fe XIV 211 \AA\ (2.0 MK), Fe XVI 335 \AA\ (2.5 MK), Fe XVIII 94 \AA\ (6.3 MK) and 
Fe VII XXI 131 \AA\ (0.4 MK, 10 MK), respectively \citep{Lemen2012}. Like most other jets in movie M1, this 
jet shows an ``anemone" (``Reverse-Y" or ``Eiffel-Tower") shape (the elongated jet body and its loop-system base), which is common among 
such solar jets \citep[e.g.,][]{Shibata2007} and can be well explained by existing jet models \citep[e.g.,][]{Canfield1996}. Easily 
figured out from the UV/EUV observations, the jet-associated loops, appearing as brightness-enhanced arches, have one root 
region at the emerging negative flux ``N1". The other footpoint is at the edge of the main positive polarity, 
indicating that the jet should be triggered by reconnection between the emerging flux and ambient field lines from the positive polarities.

Along the 13$''$-wide slit, shown as the green dashed line in Figure~\ref{jet}(b), we construct the time-distance 
diagram in Figure~\ref{slit}(a). Cool and hot components of jet materials are shown in the tricolor channels. 
It turns out to be a ``jet spectrum", which shows the jet activities causing different intensity enhancements at different passbands 
along the given slit. Figure~\ref{slit}(b) depicts the normalized integrated intensity of 8 AIA passbands over the blue dashed box 
in Figure~\ref{jet}(b). Obviously, most jet eruptions found in panel (a) correspond to distinct intensity 
enhancements in all 8 utilized passbands at the emerging negative polarity ``N1". The region in gray shadow shows the time interval 
when there is a series of violent jet eruptions (referred to as ``SJ" hereafter), during which jets are longer and more frequent.

\section{Temporal Relationship between the Emerging Flux and Jets}
\begin{figure}[tbh!]
\centering
\includegraphics[width=\hsize]{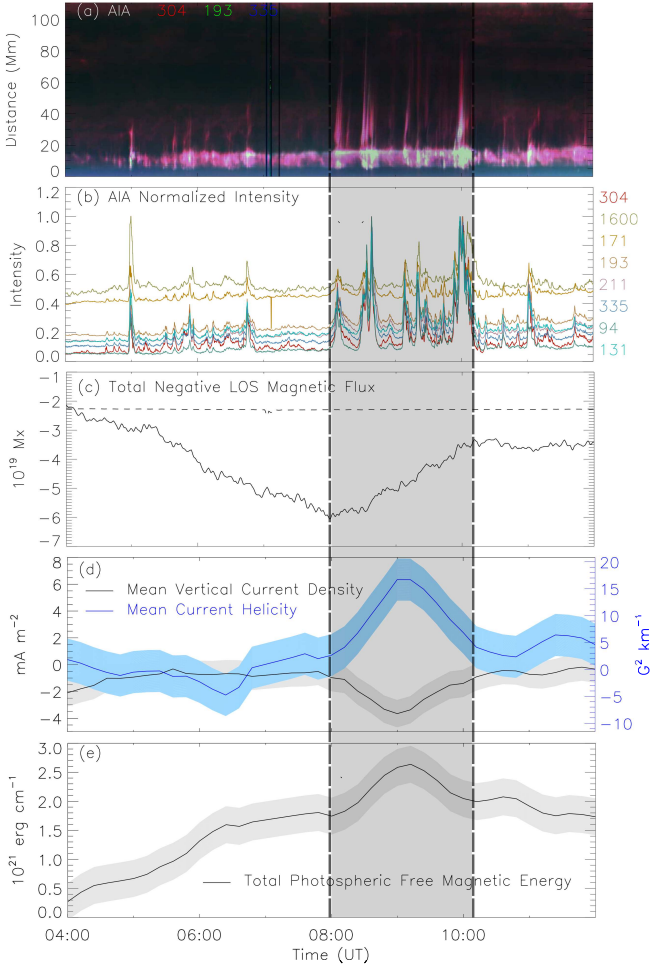}
\caption{Panel (a): Time-distance diagram of AIA 304 \AA\ (red), 193 \AA\ (green) and 335 \AA\ (blue) passbands along 
the slit in Fig.~\ref{jet}(b) from 04:00 UT to 12:00 UT. Panel (b): Normalized integrated intensities at 8 
AIA passbands within the region confined by the blue dashed box in Fig.~\ref{jet}(b). Panel (c): Total negative 
magnetic flux (solid curve) integrating the HMI LOS magnetic field data within the blue dashed box Fig.~\ref{mag}(e), 
and the millionth of the total unsigned LOS magnetic flux for the entire active region (dashed curve). 
Panel (d): Mean vertical current density (black solid curve) and mean current helicity (blue solid curve) within 
the same region as panel (c). Panel (e): Total photospheric free magnetic 
energy (black solid curve) within the same region as panel (c). 
Shadows in panel (d) and (e) indicate corresponding errors.}\label{slit}
\end{figure}

To further investigate the temporal relationship between the footpoint-region emerging flux and the erupting jets,
we visualize several characteristic physical parameters of the emerging flux in 
panel (c) to (e) in Figure~\ref{slit}, including the total negative LOS magnetic flux, the mean vertical current density, 
the mean current helicity, and the total photospheric free magnetic energy. 
The total negative LOS magnetic flux is obtained by integrating the negative data points within the region 
confined by the blue dashed box in Figure~\ref{mag}(e). The vertical current density is defined as:

\begin{equation}
\label{eq2}
\overline{J_z}\propto \frac{1}{N}\sum (\frac{\partial B_y}{\partial x} - \frac{\partial B_x}{\partial y}),
\end{equation}
where $B_x$ and $B_y$ are the tangential components of the vector magnetic field obtained from the SHARPs data. The mean current 
helicity is derived from:

\begin{equation}
\label{eq3}
\overline{H_c}\propto \frac{1}{N}\sum B_z J_z,
\end{equation}
where $B_z$ is the vertical magnetic field and $J_z$ the vertical current density obtained from Eq.~\ref{eq2} 
\citep{Leka_Barnes2003, Bobra2014}. To estimate the total photospheric free magnetic energy, we subtract the modulus of 
the potential magnetic field calculated by employing a Green Function method with unchanged vertical field component from the 
modulus of the observed magnetic field, followed by integrate the square of their difference over the blue dashed box in Figure~\ref{mag}(e) 
\citep[e.g.,][]{Gary1996, Wang1996}.

The total emerging negative LOS magnetic flux peaks at $-6.0\times10^{19}$ Mx at 08:00 UT, 
in the meantime the series of violent jet eruptions (``SJ") begins (Fig.~\ref{slit}(c)). After 08:00 UT, the total emerging negative 
flux begins to decrease and stabilizes at $\sim -3.5\times10^{19}$ Mx from 10:10 UT onwards, when is exactly the time ``SJ" ends.
The dashed curve in panel (c) stand for the millionth of the total unsigned LOS magnetic 
flux for the entire Active Region and it is found to be almost unchanged during 
the investigated period, which excludes the possibility that the temporal evolution of the total 
negative LOS flux in the emerging region is caused by the solar rotational effect.

The mean vertical current density (black solid curve in panel (d)), the mean current helicity (blue solid curve 
in panel (d)), and the total photospheric free magnetic energy (black solid curve in panel (e))
at the region of the emerging polarity ``N1" show almost the same evolution with 
each other. We find a continuous increase in these three parameters before ``SJ" begins, indicating the built-up 
of free energy and current helicity for the coming violent eruptions. After the eruption of several 
violent jets, part of the free energy is released and these parameters begin to decrease. Finally, these parameters stabilize 
from around 10:10 UT when ``SJ" ends. After that, these three parameters all become non-zero but smaller than the peak, consistent with 
that there are still few jet eruptions with smaller length and longer time lag after the ``SJ".

\section{Thermal and Kinetic Characteristics}

In order to estimate the temperature and electron number density of jets and their corresponding footpoint regions, 
we employ the Differential Emission Measure (DEM) method described in 
\cite{Hannah_Kontar2012} on the 6 AIA optically thin wavebands (171 \AA , 193 \AA , 211 \AA , 335 \AA , 94 \AA\ and 131
\AA ). The method allows a fast recovery of the DEM from solar data and can help to estimate 
of uncertainties in the solution. The response functions of these 6 AIA passbands will result in 
a DEM/EM detection in a temperature range from 0.5 to 32 MK.

\begin{figure}[tbh!]
\centering
\includegraphics[width=0.9\hsize]{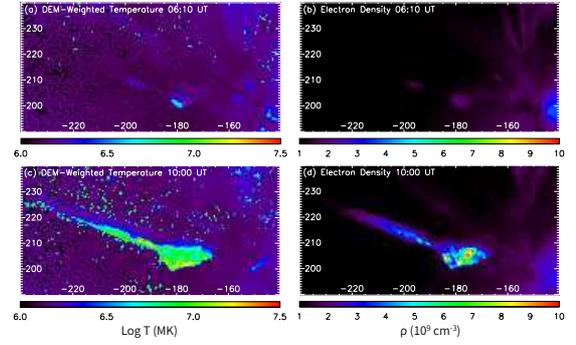}
\caption{Panel (a) and (b): Distribution of the DEM-Square-Root-Weighted temperature and the electron number density at 06:10 UT.
Panel (c) and (d): Distribution of the DEM-Square-Root-Weighted temperature and electron number density at 10:00 UT. 
Coordinates are in arcsecond.}\label{dem}
\end{figure}

Figure~\ref{dem} shows the DEM-Square-Root-Weighted (DEMSRW) temperature and electron number density distribution of 
two typical jets with one before and the other during the ``SJ" as examples. The DEMSRW temperature is defined as follows:
\begin{equation}
\label{eq4}
\overline{T}=\frac{\sum \sqrt{DEM\cdot \Delta T}\ T}{\sum\sqrt{DEM\ \Delta T}},
\end{equation}
considering that EM is proportional to the square of electron number density. The electron number density is obtained 
by $\sqrt{EM/h}$, where $EM$ is the total emission measure 
by integrating DEMs over the entire temperature range (0.5 - 32 MK), and $h$ the LOS depth which is 5 Mm, 
approximately the average width of these jets. The jet at 06:10 UT shows only one cool thread. Its most materials are 
at low-temperature ($<$ 2 MK) and low electron number density ($<$ $1.7\times 10^9$ cm$^{-3}$). On the other hand, the 
jet at 10:00 UT shows a hot thread and a relatively cool thread. The peak temperature and electron density of 
this jet are exceeding 16 MK and $7.0\times 10^9$ cm$^{-3}$, respectively.

\begin{figure}[tb!]
\centering
\includegraphics[width=0.9\hsize]{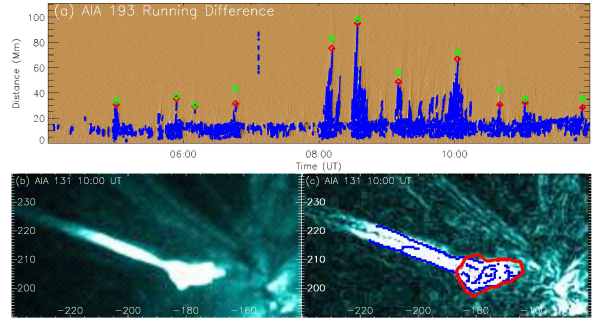}
\caption{Panel (a): Running-difference time-distance diagram along the slit in Fig.~\ref{jet}(b) at 
AIA 193 \AA\ passband. Blue dots mark the edges of jets given by the {\it Canny} edge 
detection method. Red diamonds mark the location of the jets' leading edges and green 
diamonds the location when the brightness is decreased by 50\%. 
Panel (b): AIA 131 \AA\ observation of the jet and its footpoint at 10:00 UT. Panel (c): Edge enhanced image 
of panel (b) by the {\it Sobel} edge enhancement method. Blue curves indicate the edges detected by 
the {\it Canny} algorithm and the red curve is the drawn edge of the jet's footpoint region. Coordinates are in 
units of arc second.}\label{edge}
\end{figure}

Among these jets shown in Figure~\ref{slit}(a), only few of them are found not to be associated with the emerging flux ``N1" based on 
a careful investigation of the AIA 1600 \AA\ and HMI data. After excluding them, we estimate the physical properties of 11 jets (4 before, 4 during 
and 3 after the ``SJ") and list these parameters in Table~\ref{tb1}. The second column lists the type of jets, with ``Blowout" for jets with 
several threads and ``standard" most-likely only one thread \citep[e.g.,][]{Moore2010, Morton2012}. 
3/4 before and 4/4 jets during the ``SJ" are blowout type and 3/3 after the ``SJ" are standard. To define the length of jets, we perform the {\it Sobel} edge enhancement \citep{Sobel1968} to the time-distance diagrams in AIA 304 \AA, 171 \AA, 193 \AA, 211 \AA\ and 131 \AA\ passbands, in which all the 
investigated jets cause sufficient intensity enhancements. The yellow background in Figure ~\ref{edge}(a) shows the original 
running-difference time-distance diagram in 193 \AA\ passband as an example. Then we apply the {\it Canny} edge detection algorithm \citep{Canny1986} to 
the {\it Sobel}-enhanced time-distance diagrams to define the edges of the jets (blue dots in Fig.~\ref{edge}(a)). The length of jets is then 
defined by the top of the blue dots along the trajectory of jets in the time-distance diagrams (red diamonds Fig.~\ref{edge}(a)), the corresponding 
error is determined by decreasing the brightness to half of the leading edge (green diamonds in Fig.~\ref{edge}(a)). Finally, we use the average 
value of all the lengths and errors obtained from the above five passbands for each jet. As we can see in Figure~\ref{edge}(a), the red diamonds 
always defines the leading edge with significant intensity variation and should be a lower limit of the real lengths of corresponding jets. 
The width of jets is obtained by the same method but of the original observations when jets grow to their maximum lengths (blue curves 
in Fig.~\ref{edge}(c) as an example), and the error of average width of each jet is the standard derivation of 
widths at different distances along the jet trajectory. Temperature and density are average values 
within the jet region also when it grows to its maximum length. Corresponding errors are estimated from the DEM method 
\citep{Hannah_Kontar2012}. We find that even the peak temperatures and electron number densities are rather different 
for different types of jets as shown in Figure~\ref{dem}, the average values of them do not show much diversity. The average temperature 
and electron number density for all 11 jets range from 1.2$\pm$0.3 MK to 1.5$\pm$0.3 MK, and $0.7 \pm 0.1\times 10^9 $ cm$^{-3}$ 
to $1.1 \pm 0.2 \times 10^9$ cm$^{-3}$, respectively. 

The mass of each jet is estimated by $\pi \rho L (\frac{d}{2})^2$, where $L$ is the length, $d$ the average width and $\rho$ the mass density. $\rho$ 
is set to be $\mu m_0 n_e$, where $\mu$ is the mean molecular weight 0.58 for fully ionized coronal plasma,
$m_0$ is the mass of proton and $n_e$ the estimated average electron number density of the jet. It turns out that 
the estimated mass of the jets ranges from $1.2 \pm 0.4 \times 10^{11}$ g to $38.1\pm 11.0 \times 10^{11}$ g, with the maximum mass being more than 30 times
larger than the minimum mass. We employ the equation $\frac{1}{2} m v^2$ to estimate the kinetic energy of each jet, where $m$ is the total 
mass and $v$ the projected axial speed. To obtain the projected axial speed of each jet, we apply the cross-correlation method developed by 
\cite{Tomczyk_McIntosh2009} to different parts of the time-distance diagram in Figure~\ref{slit}(a). We cross-correlate the time series at each position along the slit with the time series at the midpoint of the slit. The peak of the cross-correlation function is then fitted with a parabola such that lag or lead time at each point along the slit is returned. We then fit the lag/lead times versus the distance along the slit with a straight line - the speed (and the associated error) of the propagating jet materials is the gradient of this line. At last a comparison between the speed and the 
gradient of the inclined features in the time-distance diagram is made to check the reliability of the results. 
The projected axial speed of all 11 jets turns out to be similar and has an average value of $383\pm29$ km s$^{-1}$. 
Taking the common rotational motion along the axis for 
large-scale UV/EUV jets \citep[e.g.,][]{LiuJ2015, LiuJ2016} and the projection effect into account, we consider the kinetic energy of these jets 
estimated here to be most likely the lower limit of the true value. The kinetic energy estimated for all 11 jets ranges from 
$0.8\pm0.3\times 10^{26}$ erg to $28.0\pm9.1\times 10^{26}$ erg. Considering a fully ionized condition applicable the corona with thermal equilibrium and 
the value of Poisson constant $\gamma=5/3$, thermal energy of these jets is obtained by $3N\kappa _p T$, where $\kappa _p$ is the 
Boltzmann constant. $N=n_e V$ is the number of electrons the jet contains and $V$ is the volume of the jet. 
$T$ is the average DEMSRW temperature of the jet. Thermal energy of these jets has a range $0.6\pm0.2\times 10^{26}$ erg to $22.0\pm7.8\times 10^{26}$ erg. 
The ratios between the thermal and kinetic energy of jets are quite similar and range from 0.7 to 0.9, which is 
expected from the similar average temperature, average electron number density and axial speed for all these 11 jets.

Properties of the corresponding footpoint regions are also listed in the Table. The intensity enhancement (possibly a nano-flare) at the 
footpoint region during any of the studied jet events is not as significant as a typical flare, it is impossible to measure the strength 
of the brightening using GOES X-Ray observations or the flares' thermodynamic spectrum based on the {\it SDO}/EVE observations \citep{Wang2016}. 
An alternative option is to track 
the peak intensity of the footpoint region during each jet (which is usually several minutes before the jet gets to its 
maximum length) in the AIA 131 \AA\ passband based on the consideration that during most flares 
the intensity change of 131 \AA\ is always synchronous with that of the GOES X-Ray flux \citep[e.g.,][]{LiuK2015}. Similar as to define the 
length and width of jets, we perform the {\it Sobel} edge enhancement to the original AIA 131 \AA\ images 
(Fig.~\ref{edge}(b) at 10:00 UT as an example) to make the footpoint region clearer (Fig.~\ref{edge}(c)). 
Blue curves in Figure~\ref{edge}(c) indicate the edges detected by the {\it Canny} algorithm 
and the red curve is drawn based on considering the combination of the {\it Sobel} edge-enhanced image
and the detected {\it Canny} edges. Area and total intensity of each footpoint region then is obtained via integrating across the 
region within the red curve. A background at 03:27 UT when there is no apparent activities is subtracted 
from the obtained total intensity of each footpoint region to define the total intensity enhancement. The average temperature, average electron number density and thermal energy of each footpoint region are estimated in the same fashion with the corresponding jets.

\begin{sidewaystable}\small

\linespread{1.5} \caption{Properties of 11 Prominent Jets and Their Footpoint Regions}
\begin{center}
\begin{tabular}{c|p{1.22cm}p{1.22cm}p{1.22cm}p{1.32cm}p{1.22cm}p{1.22cm}p{1.42cm}p{1.42cm}|p{1.32cm}p{1.42cm}p{1.32cm}p{1.22cm}p{1.22cm}}
\hline
 & \multicolumn{8}{c|}{Jet} & \multicolumn{5}{c}{Footpoint Region} \\
\hline
\small
Time ($UT$)& Type & Length ($Mm$) & Width ($Mm$) & Temp ($MK$) & Density ($10^9$\ $cm^{-3}$) & Mass ($10^{11}\ g$) & Kinetic Energy ($10^{26}\ erg$) & Thermal Energy ($10^{26}\ erg$) & Area ($Mm^2$) & TIE ($10^2\ DN$) & Temp ($MK$) & Density ($10^9$\ $cm^{-3}$) & Thermal Energy ($10^{26}\ erg$) \\
\hline
05:00 & Blowout & 22.0$\pm$3.8 & 6.1$\pm$0.7 & 1.5$\pm$0.3 & 1.1$\pm$0.2 & 6.7$\pm$2.4 & 5.0$\pm$1.9 & 4.2$\pm$1.8 & 88.0$\pm$4.9 & 68.0$\pm$6.5 & 4.4$\pm$1.3 & 2.3$\pm$0.7 & 22.6$\pm$10.0 \\
05:53 & Blowout & 26.8$\pm$4.3 & 6.5$\pm$0.4 & 1.5$\pm$0.3 & 1.0$\pm$0.2 & 9.0$\pm$2.7 & 6.6$\pm$2.2 & 5.8$\pm$2.2 & 99.0$\pm$6.5 & 63.6$\pm$3.6 & 3.1$\pm$0.8 & 2.0$\pm$0.5 & 16.2$\pm$6.1 \\
06:10 & Standard & 18.7$\pm$3.2 & 2.9$\pm$0.2 & 1.2$\pm$0.3 & 1.0$\pm$0.2 & 1.2$\pm$0.4 & 0.8$\pm$0.3 & 0.6$\pm$0.2 & 47.0$\pm$3.8 & 5.7$\pm$0.5 & 2.5$\pm$0.7 & 1.7$\pm$0.4 & 2.3$\pm$0.9 \\
06:46 & Blowout & 25.6$\pm$9.9 & 7.0$\pm$0.6 & 1.3$\pm$0.3 & 0.9$\pm$0.2 & 8.4$\pm$4.0 & 6.1$\pm$3.1 & 4.7$\pm$2.4 & 108.0$\pm$6.2 & 67.0$\pm$8.9 & 4.0$\pm$1.0 & 2.0$\pm$0.5 & 24.7$\pm$9.5 \\
08:08 & Blowout & 63.0$\pm$16.2 & 4.5$\pm$0.4 & 1.4$\pm$0.3 & 0.8$\pm$0.2 & 7.9$\pm$2.9 & 5.8$\pm$2.3 & 4.6$\pm$2.0 & 126.5$\pm$5.5 & 65.5$\pm$3.3 & 3.5$\pm$1.0 & 2.1$\pm$0.6 & 17.4$\pm$7.0 \\
08:35 & Blowout & 78.6$\pm$8.7 & 9.0$\pm$0.7 & 1.4$\pm$0.3 & 0.8$\pm$0.2 & 38.1$\pm$11.0 & 28.0$\pm$9.1 & 22.0$\pm$7.8 & 129.5$\pm$7.0 & 105.7$\pm$13.0 & 4.9$\pm$1.4 & 2.9$\pm$0.8 & 68.9$\pm$28.0 \\
09:10 & Blowout & 45.1$\pm$4.6 & 4.8$\pm$0.6 & 1.3$\pm$0.3 & 0.8$\pm$0.2 & 6.5$\pm$2.1 & 4.8$\pm$1.7 & 3.7$\pm$1.4 & 104.0$\pm$5.3 & 49.8$\pm$6.8 & 3.7$\pm$1.0 & 2.2$\pm$0.6 & 17.1$\pm$7.1 \\
10:00 & Blowout & 56.3$\pm$4.3 & 7.2$\pm$0.4 & 1.4$\pm$0.3 & 0.8$\pm$0.2 & 17.7$\pm$4.5 & 13.0$\pm$3.8 & 10.8$\pm$3.6 & 102.9$\pm$8.4 & 138.1$\pm$26.3 & 5.3$\pm$1.5 & 3.4$\pm$1.0 & 54.5$\pm$22.7 \\
10:40 & Standard & 22.0$\pm$6.2 & 3.6$\pm$0.3 & 1.3$\pm$0.3 & 0.9$\pm$0.2 & 2.0$\pm$0.8 & 1.4$\pm$0.6 & 1.1$\pm$0.5 & 78.1$\pm$4.8 & 21.7$\pm$3.3 & 3.3$\pm$0.8 & 1.7$\pm$0.4 & 6.5$\pm$2.4 \\
11:04 & Standard & 23.4$\pm$5.9 & 4.2$\pm$0.4 & 1.3$\pm$0.3 & 0.9$\pm$0.2 & 2.7$\pm$1.0 & 2.0$\pm$0.8 & 1.5$\pm$0.6 & 101.5$\pm$6.0 & 53.4$\pm$8.2 & 4.8$\pm$1.3 & 2.7$\pm$0.8 & 23.2$\pm$9.5 \\
11:53 & Standard & 22.1$\pm$11.1 & 3.5$\pm$0.3 & 1.3$\pm$0.3 & 0.7$\pm$0.1 & 1.4$\pm$0.8 & 1.0$\pm$0.6 & 0.8$\pm$0.5 & 59.3$\pm$2.7 & 9.2$\pm$1.1 & 2.2$\pm$0.5 & 1.5$\pm$0.3 & 2.7$\pm$0.9 \\
\hline
\end{tabular}\\

\end{center}
\label{tb1}
Note. TIE: footpoint region Total Intensity Enhancement
\end{sidewaystable}

\begin{figure}[t!]
\centering
\includegraphics[width=1.0\hsize]{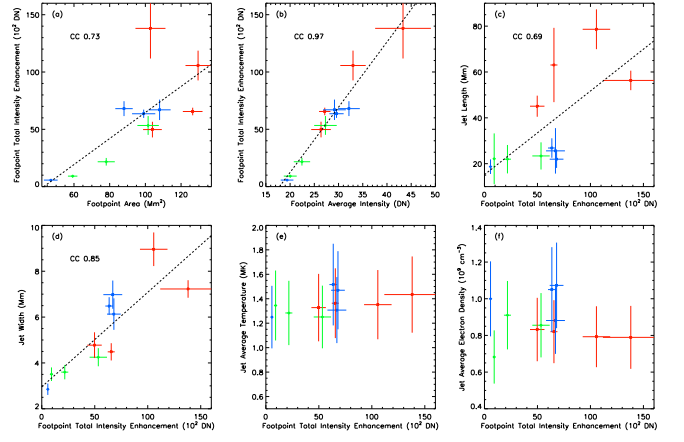}
\caption{Panel (a)-(b): The footpoint-region total 131 \AA\ intensity enhancement versus 
the footpoint-region area and footpoint-region average intensity, respectively. 
Panel (c)-(f): The jet length, average width, 
average DEMSRW temperature and average electron number density versus 
the total 131 \AA\ intensity enhancement of corresponding footpoint region. 
Red symbols mark jets during the ``SJ", blue before and 
green after the ``SJ", with diamonds for the standard and asterisks 
for the blowout type, respectively. Black dashed lines are the linear fitting results.}\label{stat}
\end{figure}

\section{Statistical Results}

Figure~\ref{stat}(a) shows the area of jet footpoint regions versus the corresponding 131 \AA\ total intensity enhancement, which presents a good positive 
relationship between them (CC $\approx$ 0.73). Red symbols mark jets during the ``SJ", blue before and green after it, with asterisks for 
blowout jets and diamonds for standard jets, respectively. Figure~\ref{stat}(b) 
illustrates that not only the area but also the intensity is positive related to the total intensity enhancement of the footpoint regions, 
suggesting that the intenser the reconnection which triggers the jet is, the larger the region it influences (the footpoint region of the jet) 
is. Most of the blowout jets have larger footpoint region and corresponding footpoint-region total intensity enhancement than standard jets, indicating 
that blowout jets should be triggered by intenser magnetic reconnection. Figure~\ref{stat}(c) to (f) show the jet length, average width, 
average temperature and average electron number density versus the total 131 \AA\ intensity enhancement of corresponding footpoint regions, respectively. 
We can see from panels (c) and (d), the jet length and width show clear positive linear relationships with the footpoint-region total intensity enhancement, 
with a relatively high cross-correlation between them (CC $\approx$ 0.69 and 0.85). As the footpoint region total intensity enhancement is considered as a 
proxy of the intensity of the magnetic reconnection, we can then conclude 
that the magnetic reconnection which triggers the jet directly influences the length and width (thus the size) of the 
jet. And, from a comparison between the diamonds and asterisks, we can see blowout jets (especially those during 
the ``SJ") tend to be longer and wider than standard jets. 
However, as shown in panel (e) and (f), the cross-correlation coefficients of the jets' average temperature and density 
with the footpoint region total intensity enhancement turn out to be very low. \cite{Shibata2007} suggest that the observed 
dominating temperature of X-ray, UV/EUV and spicule jets is mostly caused by the different heights where the jets are formed.
For the recurrent homologous jets, here, they are most likely formed at similar heights with plasmas in similar temperature and 
density - resulting in the low correlation of these two properties with the intensity of the magnetic reconnection. It is believed that 
the axial speed of a jet is comparable with the local Alfv{\'e}n speed where it is formed \citep{Shibata1992}, the similar axial speed of these jets again 
suggests that these jets should be formed at similar heights with similar magnetic field strength and plasma density.

\begin{figure}[t!]
\centering
\includegraphics[width=1.0\hsize]{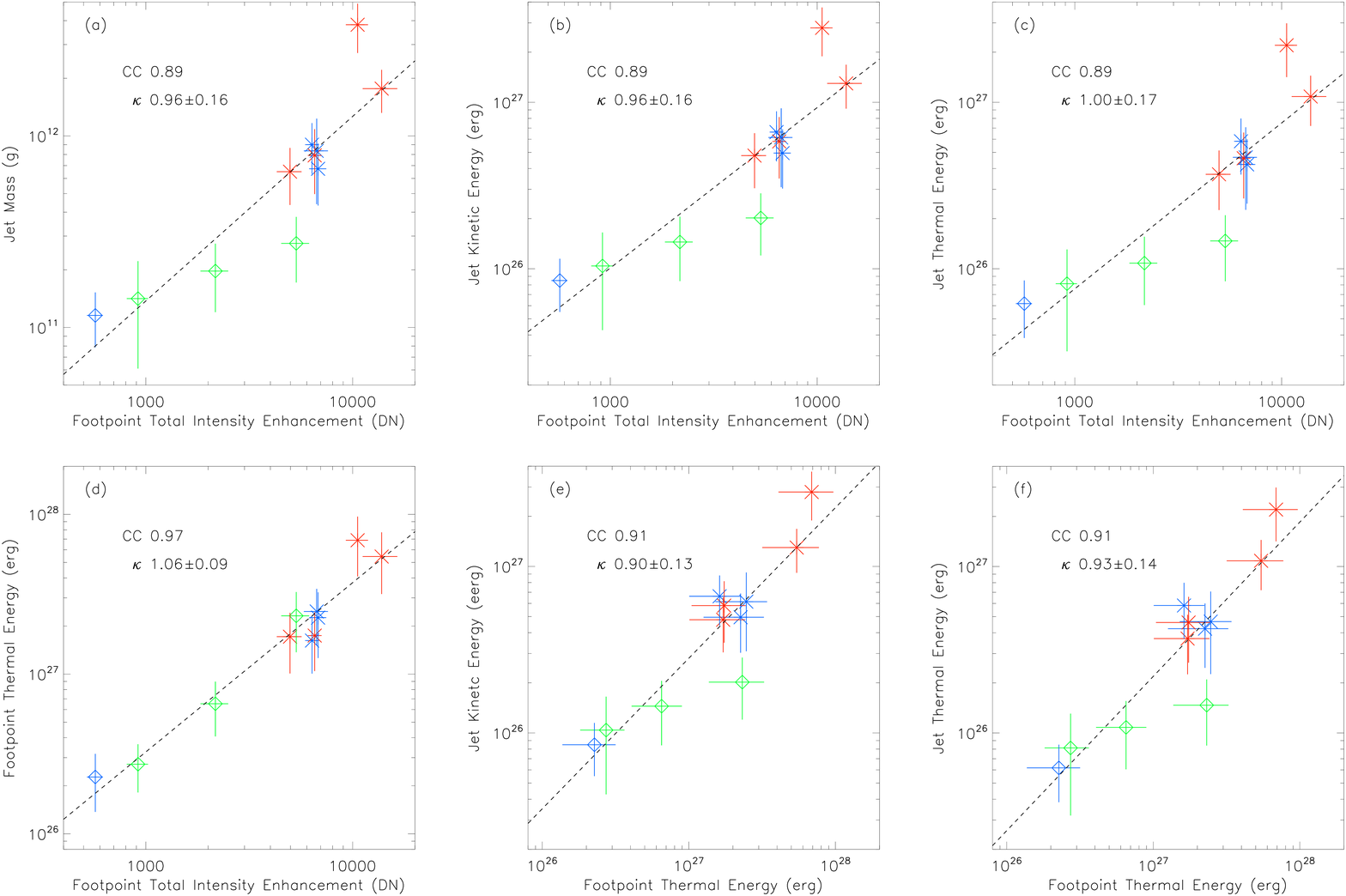}
\caption{(a)-(d): The jet mass, axial kinetic energy, thermal energy and the footpoint-region thermal energy versus 
the total 131 \AA\ intensity enhancement of corresponding footpoint region. (e)-(f): The jet kinetic energy and thermal 
energy versus the footpoint-region thermal energy. Same as Fig.~\ref{stat},
red symbols mark jets during the ``SJ", blue before and 
green after the ``SJ", with diamonds for the standard and asterisks 
for the blowout type, respectively. Black dashed lines are the linear fitting results.}\label{stat2}
\end{figure}

The relationship between the jet mass and the footpoint-region total intensity enhancement is shown in Figure~\ref{stat2}(a). They also show 
very high correlation (CC $\approx$ 0.89) with each other. Linear fit in log yields a slope of 0.96$\pm$0.16 ($\sim$1), showing the linear correlation between them. Considering the formula of the jet axial 
kinetic energy $\frac{1}{2}mv^2$ and the axial speed $v$ is similar for different jets, the jet axial kinetic energy should be 
also linearly positive related to the footpoint-region total intensity enhancement, which is proved in Figure~\ref{stat2}(b). The index is the same 
for the jet thermal energy (Fig.~\ref{stat2}(c)), as the jet thermal energy is proportional to $n_eLd^2T$ and the jet average $T$ and 
average electron number density $n_e$ are not related to the footpoint-region total intensity enhancement 
(Fig.~\ref{stat}(e) and (f)). The similar linearly positive relationship also applies to that between the total intensity enhancement 
and the thermal energy of the footpoint regions.

The total free magnetic energy released during a jet event consists of at least the following four parts: footpoint region thermal energy (heating of 
post-flare loops etc.), non-thermal energy during the nano-flare \citep{Testa2014}, jet kinetic energy 
(axial and rotational) and jet thermal energy (heating of jet materials). We are unable to estimate the non-thermal energy released 
during these jet events and most of the rotational kinetic energy is not directly resulted from the reconnection \citep{LiuJ2014}. Thus,  
we only investigate the relationship between the jet axial kinetic energy, the jet thermal energy and the footpoint-region thermal energy 
in Figure~\ref{stat2}. From Figure~\ref{stat2}(e), there is a very strong linearly positive relationship between the footpoint-region thermal 
energy and the jet axial kinetic energy (CC $\approx$ 0.91). The same linearly positive relationship 
also applies to that between the footpoint-region thermal energy 
and the jet thermal energy (CC $\approx$ 0.91, Fig.~\ref{stat2}(f)). These results demonstrate a scenario that the more heating 
reconnection contributes to the footpoint region, the more heating and work reconnection contributes to the erupted jet material. This is 
consistent with the equipartition between kinetic and thermal energies during magnetic reconnections \citep[e.g.][]{Priest2000}. Comparison 
between the diamonds and asterisks in all 6 panels in Figure~\ref{stat2} also reveals that blowout jets tend to have brighter footpoint regions, more mass, more kinetic/thermal energy than standard jets.

\section{Summary and Conclusions}

In this paper, we have performed a detailed analysis on 11 recurrent homologous jets observed by {\it SDO}/AIA 
and the related emerging (negative) flux observed by {\it SDO}/HMI from 03:00 UT to 12:00 UT on July 9th 2015. 
Let us summarize what we have learned from studying these jets and the evolution of the emerging flux as follows:

The (negative) flux ``N1" emerges at 03:05 UT and only 6 minutes after that a jet erupts. After the emergence of 
``N1", it becomes larger and larger, and more jets erupt. After 06:55 UT, the emerging flux seems 
to ``calm down" for about 1 hour. At 08:04 UT, a jet with scale much larger than that of any before pops out from this emerging flux and it 
is only the beginning of a series of frequent eruptions (``SJ"). The ``SJ" lasts for about 2 hours until 10:10 UT. 
After the ``SJ", there are still a few jets erupting from ``N1", but smaller in size. 

 
Comparing all jets erupted from 04:00 UT to 12:00 UT, we find that all jets after the
``SJ" are standard jets with only one thread, while most jets before and all jets during the ``SJ" are blowout
jets with several threads. Most of the jets show multi-thermal nature and can be observed in all 7 UV/EUV AIA 
passbands. Temporal investigation on the jets and the corresponding footpoint brightening shows that most jets 
have caused distinct intensity enhancements in all 8 utilized AIA passbands at footpoint regions.
Comparing with the total (negative) LOS magnetic flux in the footpoint region, it is found that the ``SJ" begins 
when the total (negative) flux peaks and starts to decrease. We further find that the mean vertical current density, the mean current helicity, 
and the total photospheric free magnetic energy, all 
continuously increase before and then decrease during the ``SJ". The above results suggest that with more available free magnetic 
energy, the eruptions of jets tend to be more violent, frequent and the blowout-like.

Among all the jets excluding those not related to the emerging flux, we investigate the 
properties of 11 jets and their corresponding footpoint regions. By utilizing a cross-correlation method, we find 
that all the jets have very similar projected axial speed around 383 km s$^{-1}$. 
Although the peak temperature and electron density of the blowout jets are significantly 
larger than those standard jets, the average values do not differ too much and are not related to the 
footpoint-region total 131 \AA\ intensity enhancement, indicating that these jets should be formed by materials from 
similar origins.

The length and width of the jets show strong linearly positive relationships with the corresponding footpoint-region total intensity 
enhancement, indicating that a stronger footpoint-region reconnection induces a larger jet. With a cross-correlation coefficient of about 0.9, the 
mass of jets turns out to be linearly positive relative to the footpoint-region total intensity enhancement. The kinetic and 
thermal energy of jets also show the similar relationships with the footpoint-region total intensity enhancement, indicating that 
the intenser the footpoint region reconnection is, the more energy it injects into the jet. There are also very strong 
linearly positive relationships between the footpoint-region thermal energy and the kinetic/thermal energy of jets 
(CC $\approx$ 0.9), suggesting that the more heating reconnection contributes to the footpoint region plasma, the more 
heating and work reconnection contributes to the jet material. All the above results confirm the direct relationship between  
magnetic reconnection and jets, and validate the important role of magnetic reconnection in transporting
large amount of free magnetic energy into jets.

However, as the number of homologous jets investigated in this paper is limited, it is hard to examine whether it is 
universal of the relationships found in this paper for different scales of jets in the Sun. Future work will include further 
statistical analysis of more homologous large-scale jets from different active regions, and combined studies of both large-scale and small-scale 
jets (spicules).

\acknowledgments{We acknowledge the use of data from AIA and HMI instrument
onboard Solar Dynamics Observatory ({\it SDO}). {\it SDO} is a mission for 
NASA's Living With a Star (LWS) program. We thank I.G. Hannah and E.P. Kontar for the DEM codes. J.L. acknowledges support from the China Postdoctoral Science Foundation (2015M580540). 
This work is also supported by grants from the Fundamental Research Funds for the Central Universities,
CAS (Key Research Program KZZD-EW-01-4), 
NSFC (41131065), and MOEC (20113402110001). R.E. acknowledges the support received by the Chinese Academy of Sciences President's International Fellowship Initiative, Grant No. 2016VMA045, the Science and Technology Facility Council (STFC), UK and the Royal Society (UK). R.L. acknowledges the support from the Thousand Young Talents Program of China and NSFC 41474151.}


\begin{thebibliography}{58}
\providecommand{\natexlab}[1]{#1}
\expandafter\ifx\csname urlstyle\endcsname\relax
  \providecommand{\doi}[1]{doi:\discretionary{}{}{}#1}\else
  \providecommand{\doi}{doi:\discretionary{}{}{}\begingroup
  \urlstyle{rm}\Url}\fi

\bibitem[{\textit{{Archontis} et~al.}(2010)\textit{{Archontis}, {Tsinganos},
  and {Gontikakis}}}]{Archontis2010}
{Archontis}, V., K.~{Tsinganos}, and C.~{Gontikakis}, {Recurrent solar jets in
  active regions}, \textit{\aap}, \textit{512}, L2,
  \doi{10.1051/0004-6361/200913752}, 2010.

\bibitem[{\textit{{Beckers}}(1968)}]{Becker1968}
{Beckers}, J.~M., {Solar Spicules (Invited Review Paper)}, \textit{Sol. Phys.},
  \textit{3}, 367--433, 1968.

\bibitem[{\textit{Bennett and Erd{\'e}lyi}(2015)}]{Bennett2015}
Bennett, S.~M., and R.~Erd{\'e}lyi, On the statistics of macrospicules,
  \textit{The Astrophysical Journal}, \textit{808}(2), 135, 2015.

\bibitem[{\textit{{Bobra} et~al.}(2014)\textit{{Bobra}, {Sun}, {Hoeksema},
  {Turmon}, {Liu}, {Hayashi}, {Barnes}, and {Leka}}}]{Bobra2014}
{Bobra}, M.~G., X.~{Sun}, J.~T. {Hoeksema}, M.~{Turmon}, Y.~{Liu},
  K.~{Hayashi}, G.~{Barnes}, and K.~D. {Leka}, {The Helioseismic and Magnetic
  Imager (HMI) Vector Magnetic Field Pipeline: SHARPs - Space-Weather HMI
  Active Region Patches}, \textit{\solphys}, \textit{289}, 3549--3578, 2014.

\bibitem[{\textit{{Bohlin} et~al.}(1975)\textit{{Bohlin}, {Vogel}, {Purcell},
  {Sheeley}, {Tousey}, and {Vanhoosier}}}]{Bohlin1975}
{Bohlin}, J.~D., S.~N. {Vogel}, J.~D. {Purcell}, N.~R. {Sheeley}, Jr.,
  R.~{Tousey}, and M.~E. {Vanhoosier}, {A newly observed solar feature -
  Macrospicules in He II 304 A}, \textit{\apjl}, \textit{197}, L133--L135,
  1975.

\bibitem[{\textit{{Bray} and {Loughhead}}(1964)}]{Bray1964}
{Bray}, R.~J., and R.~E. {Loughhead}, \textit{{Sunspots}}, 1964.

\bibitem[{\textit{{Canfield} et~al.}(1996)\textit{{Canfield}, {Reardon},
  {Leka}, {Shibata}, {Yokoyama}, and {Shimojo}}}]{Canfield1996}
{Canfield}, R.~C., K.~P. {Reardon}, K.~D. {Leka}, K.~{Shibata}, T.~{Yokoyama},
  and M.~{Shimojo}, {H alpha Surges and X-Ray Jets in AR 7260}, \textit{\apj},
  \textit{464}, 1016, 1996.

\bibitem[{\textit{Canny}(1986)}]{Canny1986}
Canny, J., A computational approach to edge detection, \textit{IEEE
  Transactions on Pattern Analysis and Machine Intelligence},
  \textit{PAMI-8}(6), 679--698, 1986.

\bibitem[{\textit{Chen et~al.}(2015)\textit{Chen, Su, Yin, Priya, Zhang, Liu,
  Xu, and Yu}}]{Chen2015}
Chen, J., J.~Su, Z.~Yin, T.~G. Priya, H.~Zhang, J.~Liu, H.~Xu, and S.~Yu,
  Recurrent solar jets induced by a satellite spot and moving magnetic
  features, \textit{The Astrophysical Journal}, \textit{815}(1), 71, 2015.

\bibitem[{\textit{{Cirtain} et~al.}(2007)\textit{{Cirtain}, {Golub},
  {Lundquist}, {van Ballegooijen}, {Savcheva}, {Shimojo}, {DeLuca}, {Tsuneta},
  {Sakao}, {Reeves}, {Weber}, {Kano}, {Narukage}, and
  {Shibasaki}}}]{Cirtain2007}
{Cirtain}, J.~W., L.~{Golub}, L.~{Lundquist}, A.~{van Ballegooijen},
  A.~{Savcheva}, M.~{Shimojo}, E.~{DeLuca}, S.~{Tsuneta}, T.~{Sakao},
  K.~{Reeves}, M.~{Weber}, R.~{Kano}, N.~{Narukage}, and K.~{Shibasaki},
  {Evidence for Alfv{\'e}n Waves in Solar X-ray Jets}, \textit{Sci.},
  \textit{318}, 1580, 2007.

\bibitem[{\textit{{Cranmer} and {Woolsey}}(2015)}]{Cranmer2015}
{Cranmer}, S.~R., and L.~N. {Woolsey}, {Driving Solar Spicules and Jets with
  Magnetohydrodynamic Turbulence: Testing a Persistent Idea}, \textit{\apj},
  \textit{812}, 71, 2015.

\bibitem[{\textit{{De Pontieu} et~al.}(2007)\textit{{De Pontieu}, {McIntosh},
  {Hansteen}, {Carlsson}, {Schrijver}, {Tarbell}, {Title}, {Shine}, {Suematsu},
  {Tsuneta}, {Katsukawa}, {Ichimoto}, {Shimizu}, and {Nagata}}}]{DePontieu2007}
{De Pontieu}, B., S.~{McIntosh}, V.~H. {Hansteen}, M.~{Carlsson}, C.~J.
  {Schrijver}, T.~D. {Tarbell}, A.~M. {Title}, R.~A. {Shine}, Y.~{Suematsu},
  S.~{Tsuneta}, Y.~{Katsukawa}, K.~{Ichimoto}, T.~{Shimizu}, and S.~{Nagata},
  {A Tale of Two Spicules: The Impact of Spicules on the Magnetic
  Chromosphere}, \textit{PASJ}, \textit{59}, 655, 2007.

\bibitem[{\textit{{Fang} et~al.}(2014)\textit{{Fang}, {Fan}, and
  {McIntosh}}}]{Fang2014}
{Fang}, F., Y.~{Fan}, and S.~W. {McIntosh}, {Rotating Solar Jets in Simulations
  of Flux Emergence with Thermal Conduction}, \textit{\apjl}, \textit{789},
  L19, 2014.

\bibitem[{\textit{{Gary}}(1996)}]{Gary1996}
{Gary}, G.~A., {Potential Field Extrapolation Using Three Components of a Solar
  Vector Magnetogram with a Finite Field of View}, \textit{\solphys},
  \textit{163}, 43--64, 1996.

\bibitem[{\textit{Guo et~al.}(2013)\textit{Guo, D{\'e}moulin, Schmieder, Ding,
  Dom{\'\i}nguez, and Liu}}]{Guo2013}
Guo, Y., P.~D{\'e}moulin, B.~Schmieder, M.~Ding, S.~V. Dom{\'\i}nguez, and
  Y.~Liu, Recurrent coronal jets induced by repetitively accumulated electric
  currents, \textit{Astronomy \& Astrophysics}, \textit{555}, A19, 2013.

\bibitem[{\textit{{Hannah} and {Kontar}}(2012)}]{Hannah_Kontar2012}
{Hannah}, I.~G., and E.~P. {Kontar}, {Differential emission measures from the
  regularized inversion of Hinode and SDO data}, \textit{\aap}, \textit{539},
  A146, 2012.

\bibitem[{\textit{{Kuridze} et~al.}(2015)\textit{{Kuridze}, {Henriques},
  {Mathioudakis}, {Erd{\'e}lyi}, {Zaqarashvili}, {Shelyag}, {Keys}, and
  {Keenan}}}]{Kuridze2015}
{Kuridze}, D., V.~{Henriques}, M.~{Mathioudakis}, R.~{Erd{\'e}lyi}, T.~V.
  {Zaqarashvili}, S.~{Shelyag}, P.~H. {Keys}, and F.~P. {Keenan}, {The Dynamics
  of Rapid Redshifted and Blueshifted Excursions in the Solar H{$\alpha$}
  Line}, \textit{\apj}, \textit{802}, 26, 2015.

\bibitem[{\textit{{Leka}}(1997)}]{Leka1997}
{Leka}, K.~D., {The Vector Magnetic Fields and Thermodynamics of Sunspot Light
  Bridges: The Case for Field-free Disruptions in Sunspots}, \textit{\apj},
  \textit{484}, 900--919, 1997.

\bibitem[{\textit{{Leka} and {Barnes}}(2003)}]{Leka_Barnes2003}
{Leka}, K.~D., and G.~{Barnes}, {Photospheric Magnetic Field Properties of
  Flaring versus Flare-quiet Active Regions. II. Discriminant Analysis},
  \textit{\apj}, \textit{595}, 1296--1306, 2003.

\bibitem[{\textit{{Lemen} et~al.}(2012)\textit{{Lemen}, {Title}, {Akin},
  {Boerner}, {Chou}, {Drake}, {Duncan}, {Edwards}, {Friedlaender}, {Heyman},
  {Hurlburt}, {Katz}, {Kushner}, {Levay}, {Lindgren}, {Mathur}, {McFeaters},
  {Mitchell}, {Rehse}, {Schrijver}, {Springer}, {Stern}, {Tarbell}, {Wuelser},
  {Wolfson}, {Yanari}, {Bookbinder}, {Cheimets}, {Caldwell}, {Deluca}, {Gates},
  {Golub}, {Park}, {Podgorski}, {Bush}, {Scherrer}, {Gummin}, {Smith}, {Auker},
  {Jerram}, {Pool}, {Soufli}, {Windt}, {Beardsley}, {Clapp}, {Lang}, and
  {Waltham}}}]{Lemen2012}
{Lemen}, J.~R., A.~M. {Title}, D.~J. {Akin}, P.~F. {Boerner}, C.~{Chou}, J.~F.
  {Drake}, D.~W. {Duncan}, C.~G. {Edwards}, F.~M. {Friedlaender}, G.~F.
  {Heyman}, N.~E. {Hurlburt}, N.~L. {Katz}, G.~D. {Kushner}, M.~{Levay}, R.~W.
  {Lindgren}, D.~P. {Mathur}, E.~L. {McFeaters}, S.~{Mitchell}, R.~A. {Rehse},
  C.~J. {Schrijver}, L.~A. {Springer}, R.~A. {Stern}, T.~D. {Tarbell}, J.-P.
  {Wuelser}, C.~J. {Wolfson}, C.~{Yanari}, J.~A. {Bookbinder}, P.~N.
  {Cheimets}, D.~{Caldwell}, E.~E. {Deluca}, R.~{Gates}, L.~{Golub}, S.~{Park},
  W.~A. {Podgorski}, R.~I. {Bush}, P.~H. {Scherrer}, M.~A. {Gummin},
  P.~{Smith}, G.~{Auker}, P.~{Jerram}, P.~{Pool}, R.~{Soufli}, D.~L. {Windt},
  S.~{Beardsley}, M.~{Clapp}, J.~{Lang}, and N.~{Waltham}, {The Atmospheric
  Imaging Assembly (AIA) on the Solar Dynamics Observatory (SDO)},
  \textit{\solphys}, \textit{275}, 17--40, 2012.

\bibitem[{\textit{Li et~al.}(2015)\textit{Li, Jiang, Yang, Bi, and
  Liang}}]{Li2015}
Li, H.~D., Y.~C. Jiang, J.~Y. Yang, Y.~Bi, and H.~F. Liang, The quasi-periodic
  behavior of recurrent jets caused by emerging magnetic flux,
  \textit{Astrophysics and Space Science}, \textit{359}(2), 1--6, 2015.

\bibitem[{\textit{{Liu} et~al.}(2014)\textit{{Liu}, {Wang}, {Liu}, {Zhang},
  {Liu}, {Shen}, and {Wang}}}]{LiuJ2014}
{Liu}, J., Y.~{Wang}, R.~{Liu}, Q.~{Zhang}, K.~{Liu}, C.~{Shen}, and S.~{Wang},
  {When and how does a Prominence-like Jet Gain Kinetic Energy?},
  \textit{\apj}, \textit{782}, 94, 2014.

\bibitem[{\textit{{Liu} et~al.}(2015{\natexlab{a}})\textit{{Liu}, {McIntosh},
  {De Moortel}, and {Wang}}}]{LiuJ2015b}
{Liu}, J., S.~W. {McIntosh}, I.~{De Moortel}, and Y.~{Wang}, {On the Parallel
  and Perpendicular Propagating Motions Visible inPolar Plumes: An Incubator
  For (Fast) Solar Wind Acceleration?}, \textit{\apj}, \textit{806}, 273,
  \doi{10.1088/0004-637X/806/2/273}, 2015{\natexlab{a}}.

\bibitem[{\textit{{Liu} et~al.}(2015{\natexlab{b}})\textit{{Liu}, {Wang},
  {Shen}, {Liu}, {Pan}, and {Wang}}}]{LiuJ2015}
{Liu}, J., Y.~{Wang}, C.~{Shen}, K.~{Liu}, Z.~{Pan}, and S.~{Wang}, {A Solar
  Coronal Jet Event Triggers a Coronal Mass Ejection}, \textit{\apj},
  \textit{813}, 115, 2015{\natexlab{b}}.

\bibitem[{\textit{{Liu} et~al.}(2016)\textit{{Liu}, {Fang}, {Wang}, {McIntosh},
  {Fan}, and {Zhang}}}]{LiuJ2016}
{Liu}, J., F.~{Fang}, Y.~{Wang}, S.~W. {McIntosh}, Y.~{Fan}, and Q.~{Zhang},
  {On the Observation and Simulation of Solar Coronal Twin Jets},
  \textit{\apj}, \textit{817}, 126, 2016.

\bibitem[{\textit{Liu et~al.}(2015)\textit{Liu, Wang, Zhang, Cheng, Liu, and
  Shen}}]{LiuK2015}
Liu, K., Y.~Wang, J.~Zhang, X.~Cheng, R.~Liu, and C.~Shen, Extremely large euv
  late phase of solar flares, \textit{The Astrophysical Journal},
  \textit{802}(1), 35, 2015.

\bibitem[{\textit{{Liu}}(2012)}]{LiuS2012}
{Liu}, S., {A Coronal Jet Ejection from a Sunspot Light Bridge},
  \textit{\pasa}, \textit{29}, 193--201, 2012.

\bibitem[{\textit{{Moore} et~al.}(2010)\textit{{Moore}, {Cirtain}, {Sterling},
  and {Falconer}}}]{Moore2010}
{Moore}, R.~L., J.~W. {Cirtain}, A.~C. {Sterling}, and D.~A. {Falconer},
  {Dichotomy of Solar Coronal Jets: Standard Jets and Blowout Jets},
  \textit{\apj}, \textit{720}, 757--770, 2010.

\bibitem[{\textit{Moore et~al.}(2011)\textit{Moore, Sterling, Cirtain, and
  Falconer}}]{Moore2011}
Moore, R.~L., A.~C. Sterling, J.~W. Cirtain, and D.~A. Falconer, Solar x-ray
  jets, type-ii spicules, granule-size emerging bipoles, and the genesis of the
  heliosphere, \textit{ApJ Letters}, \textit{731}(1), L18, 2011.

\bibitem[{\textit{{Moore} et~al.}(2015)\textit{{Moore}, {Sterling}, and
  {Falconer}}}]{Moore2015}
{Moore}, R.~L., A.~C. {Sterling}, and D.~A. {Falconer}, {Magnetic Untwisting in
  Solar Jets that Go into the Outer Corona in Polar Coronal Holes},
  \textit{\apj}, \textit{806}, 11, 2015.

\bibitem[{\textit{{Moreno-Insertis} et~al.}(2008)\textit{{Moreno-Insertis},
  {Galsgaard}, and {Ugarte-Urra}}}]{Moreno-Insertis2008}
{Moreno-Insertis}, F., K.~{Galsgaard}, and I.~{Ugarte-Urra}, {Jets in Coronal
  Holes: Hinode Observations and Three-dimensional Computer Modeling},
  \textit{\apjl}, \textit{673}, L211, 2008.

\bibitem[{\textit{{Morton} et~al.}(2012)\textit{{Morton}, {Srivastava}, and
  {Erd{\'e}lyi}}}]{Morton2012}
{Morton}, R.~J., A.~K. {Srivastava}, and R.~{Erd{\'e}lyi}, {Observations of
  quasi-periodic phenomena associated with a large blowout solar jet},
  \textit{\aap}, \textit{542}, A70, 2012.

\bibitem[{\textit{{Murray} et~al.}(2009)\textit{{Murray}, {van
  Driel-Gesztelyi}, and {Baker}}}]{Murray2009}
{Murray}, M.~J., L.~{van Driel-Gesztelyi}, and D.~{Baker}, {Simulations of
  emerging flux in a coronal hole: oscillatory reconnection}, \textit{\aap},
  \textit{494}, 329--337, 2009.

\bibitem[{\textit{{Pariat} et~al.}(2009)\textit{{Pariat}, {Antiochos}, and
  {DeVore}}}]{Pariat2009}
{Pariat}, E., S.~K. {Antiochos}, and C.~R. {DeVore}, {A Model for Solar Polar
  Jets}, \textit{\apj}, \textit{691}, 61--74, 2009.

\bibitem[{\textit{Pariat et~al.}(2010)\textit{Pariat, Antiochos, and
  DeVore}}]{Pariat2010}
Pariat, E., S.~K. Antiochos, and C.~R. DeVore, Three-dimensional modeling of
  quasi-homologous solar jets, \textit{The Astrophysical Journal},
  \textit{714}(2), 1762, 2010.

\bibitem[{\textit{{Pariat} et~al.}(2015)\textit{{Pariat}, {Dalmasse}, {DeVore},
  {Antiochos}, and {Karpen}}}]{Pariat2015}
{Pariat}, E., K.~{Dalmasse}, C.~R. {DeVore}, S.~K. {Antiochos}, and J.~T.
  {Karpen}, {Model for straight and helical solar jets. I. Parametric studies
  of the magnetic field geometry}, \textit{\aap}, \textit{573}, A130, 2015.

\bibitem[{\textit{{Priest} and {Forbes}}(2000)}]{Priest2000}
{Priest}, E., and T.~{Forbes} (Eds.), \textit{{Magnetic reconnection : MHD
  theory and applications}}, 2000.

\bibitem[{\textit{{Roy}}(1973)}]{Roy1973}
{Roy}, J.-R., {The Dynamics of Solar Surges}, \textit{Sol. Phys.}, \textit{32},
  139--151, 1973.

\bibitem[{\textit{{Scullion} et~al.}(2009)\textit{{Scullion}, {Popescu},
  {Banerjee}, {Doyle}, and {Erd{\'e}lyi}}}]{Scullion2009}
{Scullion}, E., M.~D. {Popescu}, D.~{Banerjee}, J.~G. {Doyle}, and
  R.~{Erd{\'e}lyi}, {Jets in Polar Coronal Holes}, \textit{\apj}, \textit{704},
  1385--1395, 2009.

\bibitem[{\textit{Shibata et~al.}(1992)\textit{Shibata, Ishido, Action, Strong,
  Hirayama, Uchida, McAllister, Matsumoto, Tsuneta, Shimizu, Hara, Sakurai,
  Ichimoto, Nishino, and Ogawara}}]{Shibata1992}
Shibata, K., Y.~Ishido, L.~W. Action, K.~T. Strong, T.~Hirayama, Y.~Uchida,
  A.~H. McAllister, R.~Matsumoto, S.~Tsuneta, T.~Shimizu, H.~Hara, T.~Sakurai,
  K.~Ichimoto, Y.~Nishino, and Y.~Ogawara, {Observations of X-Ray Jets with the
  Yohkoh Soft X-Ray Telescope}, \textit{PASJ}, \textit{44}, L173--L179, 1992.

\bibitem[{\textit{{Shibata} et~al.}(1996)\textit{{Shibata}, {Shimojo},
  {Yokoyama}, and {Ohyama}}}]{Shibata1996}
{Shibata}, K., M.~{Shimojo}, T.~{Yokoyama}, and M.~{Ohyama}, {Theory and
  observations of X-ray jets.}, in \textit{Astronomical Society of the Pacific
  Conference Series}, \textit{Astronomical Society of the Pacific Conference
  Series}, vol. 111, edited by R.~D. {Bentley} and J.~T. {Mariska}, pp. 29--38,
  1996.

\bibitem[{\textit{{Shibata} et~al.}(2007)\textit{{Shibata}, {Nakamura},
  {Matsumoto}, {Otsuji}, {Okamoto}, {Nishizuka}, {Kawate}, {Watanabe},
  {Nagata}, {UeNo}, {Kitai}, {Nozawa}, {Tsuneta}, {Suematsu}, {Ichimoto},
  {Shimizu}, {Katsukawa}, {Tarbell}, {Berger}, {Lites}, {Shine}, and
  {Title}}}]{Shibata2007}
{Shibata}, K., T.~{Nakamura}, T.~{Matsumoto}, K.~{Otsuji}, T.~J. {Okamoto},
  N.~{Nishizuka}, T.~{Kawate}, H.~{Watanabe}, S.~{Nagata}, S.~{UeNo},
  R.~{Kitai}, S.~{Nozawa}, S.~{Tsuneta}, Y.~{Suematsu}, K.~{Ichimoto},
  T.~{Shimizu}, Y.~{Katsukawa}, T.~D. {Tarbell}, T.~E. {Berger}, B.~W. {Lites},
  R.~A. {Shine}, and A.~M. {Title}, {Chromospheric Anemone Jets as Evidence of
  Ubiquitous Reconnection}, \textit{Science}, \textit{318}, 1591, 2007.

\bibitem[{\textit{{Sobel} and {Feldman}}(1968)}]{Sobel1968}
{Sobel}, I., and G.~{Feldman}, Presentation at the stanford artiﬁcial
  intelligence project, \textit{IEEE Transactions on Pattern Analysis and
  Machine Intelligence}, 1968.

\bibitem[{\textit{{Sterling}}(2000)}]{Sterling2000}
{Sterling}, A.~C., {Solar Spicules: A Review of Recent Models and Targets for
  Future Observations - (Invited Review)}, \textit{Sol. Phys.}, \textit{196},
  79--111, 2000.

\bibitem[{\textit{{Testa} et~al.}(2014)\textit{{Testa}, {De Pontieu}, {Allred},
  {Carlsson}, {Reale}, {Daw}, {Hansteen}, {Martinez-Sykora}, {Liu}, {DeLuca},
  {Golub}, {McKillop}, {Reeves}, {Saar}, {Tian}, {Lemen}, {Title}, {Boerner},
  {Hurlburt}, {Tarbell}, {Wuelser}, {Kleint}, {Kankelborg}, and
  {Jaeggli}}}]{Testa2014}
{Testa}, P., B.~{De Pontieu}, J.~{Allred}, M.~{Carlsson}, F.~{Reale}, A.~{Daw},
  V.~{Hansteen}, J.~{Martinez-Sykora}, W.~{Liu}, E.~E. {DeLuca}, L.~{Golub},
  S.~{McKillop}, K.~{Reeves}, S.~{Saar}, H.~{Tian}, J.~{Lemen}, A.~{Title},
  P.~{Boerner}, N.~{Hurlburt}, T.~D. {Tarbell}, J.~P. {Wuelser}, L.~{Kleint},
  C.~{Kankelborg}, and S.~{Jaeggli}, {Evidence of nonthermal particles in
  coronal loops heated impulsively by nanoflares}, \textit{Science},
  \textit{346}, 1255724, 2014.

\bibitem[{\textit{{Threlfall} et~al.}(2013)\textit{{Threlfall}, {De Moortel},
  {McIntosh}, and {Bethge}}}]{Threlfall2013}
{Threlfall}, J., I.~{De Moortel}, S.~W. {McIntosh}, and C.~{Bethge}, {First
  comparison of wave observations from CoMP and AIA/SDO}, \textit{A\&A},
  \textit{556}, A124, 2013.

\bibitem[{\textit{Tian et~al.}(2014)\textit{Tian, DeLuca, Cranmer, {De
  Pontieu}, Peter, Martinez-Sykora, Golub, McKillop, Reeves, Miralles,
  McCauley, Saar, Testa, Weber, Murphy, Lemen, Title, Boerner, Hurlburt,
  Tarbell, Wuelser, Kleint, Kankelborg, Jaeggli, Carlsson, Hansteen, and
  McIntosh}}]{Tian2014}
Tian, H., E.~E. DeLuca, S.~R. Cranmer, B.~{De Pontieu}, H.~Peter,
  J.~Martinez-Sykora, L.~Golub, S.~McKillop, K.~K. Reeves, M.~P. Miralles,
  P.~McCauley, S.~Saar, P.~Testa, M.~Weber, N.~Murphy, J.~Lemen, a.~Title,
  P.~Boerner, N.~Hurlburt, T.~D. Tarbell, J.~P. Wuelser, L.~Kleint,
  C.~Kankelborg, S.~Jaeggli, M.~Carlsson, V.~Hansteen, and S.~W. McIntosh,
  {Prevalence of small-scale jets from the networks of the solar transition
  region and chromosphere}, \textit{Sci.}, \textit{346}(6207),
  1255,711--1255,711, 2014.

\bibitem[{\textit{{Tomczyk} and {McIntosh}}(2009)}]{Tomczyk_McIntosh2009}
{Tomczyk}, S., and S.~W. {McIntosh}, {Time-Distance Seismology of the Solar
  Corona with CoMP}, \textit{\apj}, \textit{697}, 1384--1391, 2009.

\bibitem[{\textit{Toriumi et~al.}(2015)\textit{Toriumi, Cheung, and
  Katsukawa}}]{Toriumi2015}
Toriumi, S., M.~C.~M. Cheung, and Y.~Katsukawa, Light bridge in a developing
  active region. ii. numerical simulation of flux emergence and light bridge
  formation, \textit{The Astrophysical Journal}, \textit{811}(2), 138, 2015.

\bibitem[{\textit{van~der Voort et~al.}(2009)\textit{van~der Voort, Leenaarts,
  de~Pontieu, Carlsson, and Vissers}}]{vanderVoort2009}
van~der Voort, L.~R., J.~Leenaarts, B.~de~Pontieu, M.~Carlsson, and G.~Vissers,
  On-disk counterparts of type ii spicules in the ca ii 854.2 nm and hα lines,
  \textit{The Astrophysical Journal}, \textit{705}(1), 272, 2009.

\bibitem[{\textit{{Verwichte} et~al.}(2009)\textit{{Verwichte}, {Aschwanden},
  {Van Doorsselaere}, {Foullon}, and {Nakariakov}}}]{Verwichte2009}
{Verwichte}, E., M.~J. {Aschwanden}, T.~{Van Doorsselaere}, C.~{Foullon}, and
  V.~M. {Nakariakov}, {Seismology of a Large Solar Coronal Loop from
  EUVI/STEREO Observations of its Transverse Oscillation}, \textit{ApJ},
  \textit{698}, 397--404, 2009.

\bibitem[{\textit{{Wang} et~al.}(1996)\textit{{Wang}, {Shi}, {Wang}, and
  {Lue}}}]{Wang1996}
{Wang}, J., Z.~{Shi}, H.~{Wang}, and Y.~{Lue}, {Flares and the Magnetic
  Nonpotentiality}, \textit{\apj}, \textit{456}, 861, 1996.

\bibitem[{\textit{{Wang} et~al.}(2016)\textit{{Wang}, {Zhou}, {Zhang}, {Liu},
  {Liu}, {Shen}, and {Chamberlin}}}]{Wang2016}
{Wang}, Y., Z.~{Zhou}, J.~{Zhang}, K.~{Liu}, R.~{Liu}, C.~{Shen}, and P.~C.
  {Chamberlin}, {Thermodynamic Spectrum of Solar Flares Based on SDO/EVE
  Observations: Techniques and First Results}, \textit{APJS}, \textit{223}, 4,
  2016.

\bibitem[{\textit{Yang et~al.}(2013)\textit{Yang, He, Peter, Tu, Zhang, Feng,
  and Zhang}}]{Yang2013}
Yang, L., J.~He, H.~Peter, C.~Tu, L.~Zhang, X.~Feng, and S.~Zhang, Numerical
  simulations of chromospheric anemone jets associated with moving magnetic
  features, \textit{The Astrophysical Journal}, \textit{777}(1), 16, 2013.

\bibitem[{\textit{{Zhang} and {Ji}}(2014)}]{Zhang2014}
{Zhang}, Q.~M., and H.~S. {Ji}, {Blobs in recurring extreme-ultraviolet jets},
  \textit{\aap}, \textit{567}, A11, 2014.

\bibitem[{\textit{Zheng et~al.}(2016)\textit{Zheng, Chen, Du, and
  Li}}]{Zheng2016}
Zheng, R., Y.~Chen, G.~Du, and C.~Li, Solar jet–coronal hole collision and a
  closely related coronal mass ejection, \textit{The Astrophysical Journal
  Letters}, \textit{819}(2), L18, 2016.

\end{thebibliography}
\end{document}